\documentclass[12pt]{article}
\catcode`\@=11

\@addtoreset{equation}{section}
\def\theequation{\arabic{section}.\arabic{equation}}
     
\catcode`\@=11
\def\thesection{\arabic{section}.}

\def\appendix{\setcounter{section}{0}
        \def\thesection{Appendix.}
        \def\theequation{\Alph{section}.\arabic{equation}}}
\def\section{\@startsection{section}{1}{\z@}{3.5ex plus 1ex minus
   .2ex}{2.3ex plus .2ex}{\large\bf}}
     
\def\eqnarray{\let\@currentlabel=\theequation\refstepcounter{equation}
    \global\@eqnswtrue
    \global\@eqcnt\z@\tabskip\@centering\let\\=\@eqncr
    $$\halign to \displaywidth\bgroup\@eqnsel\hskip\@centering
      $\displaystyle\tabskip\z@{##}$&\global\@eqcnt\@ne 
       \hfil${{}##{}}$\hfil
      &\global\@eqcnt\tw@ $\displaystyle\tabskip\z@{##}$\hfil 
       \tabskip\@centering&\llap{##}\tabskip\z@\cr}
\def\lefteqn#1{\hbox to 4\arraycolsep{$\displaystyle #1$\hss}}
\long\def\@makefntext#1{\parindent 0cm\noindent
\hbox to 1em{\hss$^{\@thefnmark}$}#1}

\def\rref#1{(\ref{#1})}
\newcommand{\beq}{\begin{equation}}
\newcommand{\eeq}{\end{equation}}

\begin{document}

%
%
%
%
\def\citen#1{%
\edef\@tempa{\@ignspaftercomma,#1, \@end, }
\edef\@tempa{\expandafter\@ignendcommas\@tempa\@end}%
\if@filesw \immediate \write \@auxout {\string \citation {\@tempa}}\fi
\@tempcntb\m@ne \let\@h@ld\relax \let\@citea\@empty
\@for \@citeb:=\@tempa\do {\@cmpresscites}%
\@h@ld}
%
\def\@ignspaftercomma#1, {\ifx\@end#1\@empty\else
   #1,\expandafter\@ignspaftercomma\fi}
\def\@ignendcommas,#1,\@end{#1}
%
%
\def\@cmpresscites{%
 \expandafter\let \expandafter\@B@citeB \csname b@\@citeb \endcsname
 \ifx\@B@citeB\relax 
    \@h@ld\@citea\@tempcntb\m@ne{\bf ?}%
    \@warning {Citation `\@citeb ' on page \thepage \space undefined}%
 \else
    \@tempcnta\@tempcntb \advance\@tempcnta\@ne
    \setbox\z@\hbox\bgroup 
    \ifnum\z@<0\@B@citeB \relax
       \egroup \@tempcntb\@B@citeB \relax
       \else \egroup \@tempcntb\m@ne \fi
    \ifnum\@tempcnta=\@tempcntb 
       \ifx\@h@ld\relax 
          \edef \@h@ld{\@citea\@B@citeB}%
       \else 
          \edef\@h@ld{\hbox{--}\penalty\@highpenalty \@B@citeB}%
       \fi
    \else   
       \@h@ld \@citea \@B@citeB \let\@h@ld\relax
 \fi\fi%
 \let\@citea\@citepunct
}
%
\def\@citepunct{,\penalty\@highpenalty\hskip.13em plus.1em minus.1em}%
%
%
\def\@citex[#1]#2{\@cite{\citen{#2}}{#1}}%
%
%
\def\@cite#1#2{\leavevmode\unskip
  \ifnum\lastpenalty=\z@ \penalty\@highpenalty \fi 
  \ [{\multiply\@highpenalty 3 #1
      \if@tempswa,\penalty\@highpenalty\ #2\fi 
    }]\spacefactor\@m}
\let\nocitecount\relax  
%
\begin{titlepage}
\vspace{.5in}
\begin{flushright}
UCD-02-02\\
February 2002\\
gr-qc/0203001\\
\end{flushright}
\vspace{.5in}
\begin{center}
{\Large\bf
 Near-Horizon Conformal Symmetry\\[.6ex] and Black Hole Entropy}\\
\vspace{.4in}
{S.~C{\sc arlip}\footnote{\it email: carlip@dirac.ucdavis.edu}\\
       {\small\it Department of Physics}\\
       {\small\it University of California}\\
       {\small\it Davis, CA 95616}\\{\small\it USA}}
\end{center}

\vspace{.5in}
\begin{center}
{\large\bf Abstract}
\end{center}
\begin{center}
\begin{minipage}{4.5in}
{\small
Near an event horizon, the action of general relativity acquires 
a new asymptotic conformal symmetry.  Using two-dimensional 
dilaton gravity as a test case, I show that this symmetry results in 
a chiral Virasoro algebra with a calculable classical central charge, 
and that Cardy's formula for the density of states reproduces the 
Bekenstein-Hawking entropy.  This result lends support to the 
notion that the universal nature of black hole entropy is controlled 
by conformal symmetry near the horizon.
}
\end{minipage}
\end{center}
\end{titlepage}
\addtocounter{footnote}{-1}

\section{Introduction}

Since the seminal work of Bekenstein \cite{Bekenstein} and Hawking
\cite{Hawking} in the early 1970s, we have  understood that black holes 
are thermodynamic objects, with characteristic temperatures and 
entropies.  The Bekenstein-Hawking entropy depends on both Planck's
constant $\hbar$ and Newton's gravitational constant $G$, and offers
one of the few known ``windows'' into quantum gravity.  In particular,  
an understanding of the microscopic statistical mechanics of black 
hole thermodynamics may give us valuable information about the
fundamental degrees of freedom of quantized general relativity.  Until 
quite recently, though, standard derivations of the Bekenstein-Hawking 
entropy involved only macroscopic thermodynamics, and a statistical 
mechanical description was more a hope than a reality.

In the past few years, this situation has changed dramatically.  Today,  
indeed, we face the opposite problem: we have many candidate 
descriptions of black hole statistical mechanics, all of which yield the 
same entropy despite counting very different states.  In particular, there 
are two string theoretical descriptions, one based on counting D-brane 
states \cite{StromVafa} and another involving a dual conformal field theory 
\cite{Maldacena}; an approach in loop quantum gravity that counts 
spin network states \cite{Ashtekar}; and a slightly more obscure method 
\cite{Frolov} based on Sakharov's old idea of induced gravity \cite{Sakharov}.  
The problem of ``universality'' is to explain why these approaches agree, 
and why they agree with the original semiclassical computations 
\cite{Hawking,Gibbons} that know nothing of the details of quantum gravity.

One possible answer is that black hole thermodynamics may be controlled 
by a symmetry inherited from the classical theory.  This idea has its
roots in an observation by Strominger \cite{Strominger} and Birmingham
et al.\ \cite{Birmingham} that black hole entropy in three spacetime
dimensions can be obtained from Cardy's formula \cite{Cardy} for
the density of states of a two-dimensional conformal field theory at the 
``boundary'' of spacetime.  A number of authors have tried to extend such
arguments to black holes in arbitrary dimensions \cite{Carlip,Carlip2,%
Solodukhin,Brustein,Das,Jing,Cadoni,Cadoni2,Navarro,Lin,Carlip3,Park0}, 
but while these calculations seem to have the right ``flavor,'' none is yet fully 
satisfactory \cite{Park,Park2,Carlip4,Dreyer,Koga}.  In particular, all such 
proposals so far require awkward boundary conditions at black hole horizons,
and most have serious difficulties in two spacetime dimensions, where there 
does not seem to be enough room at the one-dimensional horizon for the
required degrees of freedom.
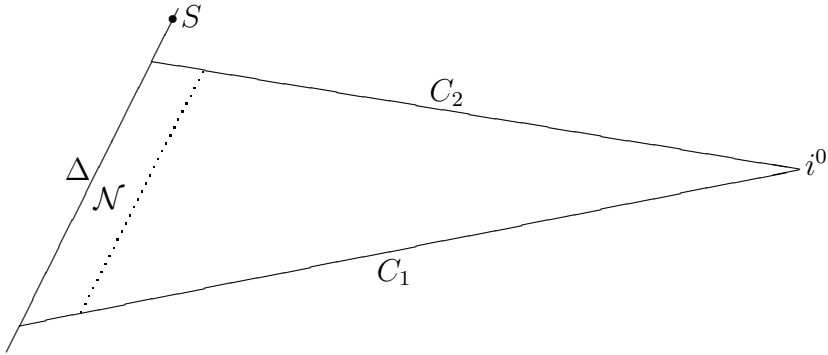
\begin{figure}
\begin{center} 
\begin{picture}(100,130)(140,20)
\put(50,20){\line(1,2){65}}
\qbezier[30](78,35)(101,81)(124,126)
\put(72,85){$\Delta$}
\put(83,74){$\cal N$}
\put(55,30){\line(5,1){295}}
\put(105,130){\line(6,-1){245}}
\put(190,47){$C_1$}
\put(210,115){$C_2$}
\put(352,87){$i^0$}
\put(113,146){\circle*{3}}
\put(116,143){$S$}
\end{picture}
\caption{A black hole spacetime: horizon $\Delta$, two partial Cauchy surfaces
$C_1$ and $C_2$, a ``reference'' cross-section $S$ of $\Delta$, and a neighborhood
$\cal N$ of the horizon.}
\end{center}
\end{figure}

In this paper, I point out three new ingredients that lead to an improved
description of the near-horizon symmetries of a black hole, and show
how they may overcome these difficulties.  The new ingredients are the
following:
\begin{enumerate}
\item {\bf Conformal symmetry:} In the presence of a stationary 
black hole---or, more generally, a black hole with a momentarily stationary 
region near its horizon---the Einstein-Hilbert action of general relativity 
acquires a new conformal symmetry.  Indeed, let $\Delta$ be a segment 
of such a horizon (see figure 1), and let $\cal N$ be a ``momentarily stationary'' 
neighborhood, that is, a neighborhood admitting a Killing vector $\chi^a$ 
for which $\Delta$ is a Killing horizon.  If $f$ is an arbitrary smooth function 
that vanishes outside $\cal N$, then under the transformation
\beq
g_{ab}\rightarrow \nabla_c(f\chi^c)g_{ab}
\label{a1}
\eeq
the action in $n$ dimensions transforms, up to possible boundary terms, as
\beq
\delta I = {1\over16\pi G}\int_N \nabla_c(f\chi^c)g^{ab}G_{ab}\epsilon
   = {1\over16\pi G}{n-2\over2}\int_N f \chi^c\nabla_c R \epsilon = 0 ,
\label{a2}
\eeq
where $\epsilon$ is the volume element and the last equality follows from 
the fact that $\chi^a$ is a Killing vector.  The addition of matter to the 
Lagrangian will not change this result, as long as the matter fields have 
the same symmetries as the metric in $\cal N$.

\hspace*{1em} For \rref{a1} to be a genuine symmetry, of course, it must 
preserve the relevant space of fields; that is, the new metric must also 
admit a Killing vector in $\cal N$.  It is straightforward to check that this will 
be the case if
\beq
(\chi^a\nabla_a)^2 f = 0 .
\label{a2a}
\eeq
Below, we shall generalize this argument to the case of an asymptotic 
symmetry, for which $(\chi^a\nabla_a)^2 f(x)$ approaches zero as $x$
approaches $\Delta$.

\hspace*{1em} The transformation \rref{a1} is not a symmetry of the full 
Einstein-Hilbert action, of course, since a generic metric admits no local 
Killing vector.  If one is interested in quantum gravitational questions about 
black holes, though, one should restrict the action to field configurations 
in which a black hole is present \cite{Carlipx}.  For such configurations, a 
symmetry of the form \rref{a1} is present at least as an asymptotic symmetry.

\item {\bf Horizon symplectic form:} In the presence of a horizon, the 
canonical symplectic form of general relativity---that is, roughly, the Poisson 
brackets---picks up a new contribution from the horizon.  This is most easily 
seen in the covariant canonical formalism \cite{Wald}, in which the symplectic 
form for a collection of fields $\phi$ is given by an integral
\beq
\Omega[\phi;\delta_1\phi,\delta_2\phi] 
   = \int_C\omega[\phi;\delta_1\phi,\delta_2\phi]
\label{a3}
\eeq
of a closed form $\omega$ over a (partial) Cauchy surface $C$.  Consider
the two surfaces $C_1$ and $C_2$ of figure 1.  The fact that $\omega$ is a
closed form ensures that
\beq
\Omega_{C_1}[\phi;\delta_1\phi,\delta_2\phi] 
   = \Omega_{C_2}[\phi;\delta_1\phi,\delta_2\phi] + 
   \int_{\Delta\cap C_1}^{\Delta\cap C_2}\!\!\omega[\phi;\delta_1\phi,\delta_2\phi] ,
\label{a4}
\eeq
where the integral on the right-hand side is over the portion of the
horizon joining $C_1$ and $C_2$.  For the isolated horizon boundary 
conditions of Ref.\ \cite{Ashtekar2}, the restriction of $\omega$ to $\Delta$
is exact, and the horizon integral can be absorbed in $\Omega$.  In general,
though, there is no reason to expect such a simple outcome.  Instead, to define 
a symplectic structure that is independent of the choice of Cauchy surface $C$,
one must choose a ``reference'' cross-section $S$ of the horizon and define
\beq
{\hat\Omega}_{C}[\phi;\delta_1\phi,\delta_2\phi] =
   \int_C\omega[\phi;\delta_1\phi,\delta_2\phi]
   + \int_S^{\Delta\cap C}\!\!\omega[\phi;\delta_1\phi,\delta_2\phi] 
\label{a5}
\eeq
where the second integral is over the portion of the horizon connecting $S$
and $C_1$.  This term is already implicit in \cite{Ashtekar2}, where the
boundary contribution to $\Omega$ is fixed in terms of a reference
cross-section that is used to determine the relevant ``integration constant.''

\hspace*{1em} The Poisson brackets thus include a contribution from the 
horizon itself.  As we shall see below, for an asymptotic symmetry of the 
sort we are interested in here, this horizon contribution will dominate.

\item {\bf Asymptotic symmetry:} The horizon of a generic black 
hole need not have a stationary neighborhood $\cal N$.  The nonexpanding 
horizon boundary conditions of Ashtekar et al.\ \cite{Ashtekar2}, for
example, require a Killing vector only on the horizon itself.  What we really 
need is the notion of an asymptotic symmetry, in which the spacetime is 
``almost'' stationary as one approaches the horizon.  

\hspace*{1em} Traditionally, an ``asymptotic symmetry'' in general 
relativity has meant an exact symmetry, i.e., a diffeomorphism, that 
preserves some extra asymptotic structure.  Here we have a slightly 
different situation: a symmetry of the action that may be exact only at 
the horizon, but that can be made arbitrarily good by shrinking the 
neighborhood $\cal N$ in which the parameter $f$ has its support.  
This circumstance is probably best viewed as an instance of a weakly 
broken symmetry.  In particular, we can find an approximate Killing 
vector $\chi^a$ near the horizon (e.g., in the manner of \cite{Matzner}) 
and a metric $\bar g$ for which $\chi^a$ is an exact Killing vector, and 
write $g = {\bar g} +  h$, where $h=0$ at the horizon.  The Lagrangian 
${\bf L}[{\bar g} + h]$ is then invariant up to terms of order $h$, and it 
may  be shown that the would-be Noether current for the transformation 
\rref{a1} is conserved up to terms of order $h$.   While more work is 
required to fully understand this sort of symmetry, it is reasonably clear 
that if $h$ is sufficiently smooth, an asymptotic symmetry near the 
horizon should become an exact symmetry for fields located on the 
horizon itself

\end{enumerate}

\section{The two-dimensional black hole}

We can now ask whether the new symmetry \rref{a1} places any restrictions
on black hole thermodynamics.  In general, one ought not expect a symmetry
to determine anything as ``microscopic'' as a density of states.  There is one 
important exception, though: for a one- or two-dimensional conformal 
symmetry described by a Virasoro algebra 
\beq
\left[L_m,L_n\right] =
     (m-n)L_{m+n} + {c\over12}m(m^2-1)\delta_{m+n,0}
\label{b1}
\eeq
with central charge $c$, the Cardy formula \cite{Cardy,Carlipy} tells us that 
the number of states having eigenvalue $\Delta$ of $L_0$ goes asymptotically 
as
\beq
\rho(\Delta) \sim
   \exp\left\{ 2\pi\sqrt{c_{\hbox{\scriptsize eff}}\Delta\over6}\right\} 
\quad \hbox{with $c_{\hbox{\scriptsize eff}} = c - 24\Delta_0$} ,
\label{b2}
\eeq
where $\Delta_0$ is the lowest eigenvalue of $L_0$.  The question is thus
whether the symmetry \rref{a1} can be described by such an algebra.

To answer this question, it is useful to focus on a particular example,
two-dimensional dilaton gravity.  This is not as restrictive as it may seem,
since general relativity in any dimension can be dimensionally reduced 
via a Kaluza-Klein mechanism to two-dimensional gravity coupled to
``matter'' fields, and I shall argue below that the extra fields do not affect
the conclusions.  Still, this work should be considered a first step, which
can presumably be considerably generalized.

The action for dilaton gravity can be written in the form \cite{Kunstatter}
\beq
I = \int{\bf L} 
   = {1\over2G}\int \left( \phi R + {1\over\ell^2}V[\phi] \right)\epsilon ,
\label{b4}
\eeq
where $\epsilon$ is the two-dimensional volume form and $V$ is an 
arbitrary function of the dilaton field $\phi$.  (The kinetic term for $\phi$ 
has been absorbed into $\phi R$ by field redefinition.)  Strictly speaking, 
one cannot define the expansion of a null congruence in two dimensions,
but the analog in dilaton gravity is
\beq
\vartheta = {1\over\phi}\ell^a\nabla_a\phi
\label{b5}
\eeq
where $\ell^a$ is the null normal.  All known exact black hole solutions,
including dimensionally reduced descriptions of higher-dimensional black
holes, have null horizons with vanishing $\vartheta$.

As in previous work \cite{Carlip,Carlip2}, we will start with a ``stretched
horizon,'' in this case a null surface $\tilde\Delta$ with null normal 
$\ell^a$ for which $\vartheta$ is small but nonzero.  Near a genuine
horizon, we can take $\vartheta$ to be a measure of how far we have 
``stretched'' away; in the end, we will take the limit $\vartheta\rightarrow0$.

In two dimensions, the vector $\ell^a$ determines a unique ``orthogonal'' 
null vector $n^a$, such that $\ell^an_a=-1$.  We extend $n^a$ from 
$\tilde\Delta$ by requiring that
\beq
n^a\nabla_a n_b = 0 ,
\label{b5a}
\eeq
from which it follows that
\beq
\nabla_a\ell_b = -\kappa n_a\ell_b , \quad
\nabla_a n_b = \kappa n_an_b
\label{b6}
\eeq
where $\kappa$ is the ``surface gravity.''  Note, though, that unlike a 
timelike or spacelike unit vector, a null normal does not have a fixed 
normalization:  by rescaling $\ell^a\rightarrow f\ell^a$, one can change 
$\kappa$ almost arbitrarily on a fixed null surface $\tilde\Delta$
\cite{Ashtekar2},
\beq
\kappa \rightarrow \ell^a\nabla_a f + \kappa f , \qquad n^a\nabla_a f = 0 ,
\label{b6a}
\eeq
where the last condition ensures that \rref{b5a} is preserved.  We shall 
use this freedom below to choose a convenient form for $\kappa$.

Observe from \rref{b6} that
\beq
\nabla_a\ell_b + \nabla_b\ell_a = \kappa g_{ab} ,
\label{b6b}
\eeq
so $\ell_a$ is a conformal Killing vector.  We will see later that the
natural scaling of $\ell_a$ leads to a surface gravity $\kappa$ 
proportional to $\vartheta$, so $\ell_a$ is actually an approximate 
Killing vector near the horizon.

The application of the transformation \rref{a1} to two dimensions 
is a bit tricky, both because a new field $\phi$ is present and
because the field equations of dilaton gravity differ from those of 
ordinary general relativity.  In general, we should expect $\phi$ as 
well as $g_{ab}$ to transform, and it is easy to check that under a 
transformation 
\begin{eqnarray}
\delta g_{ab}&=& \nabla_c(f\ell^c)g_{ab} 
     = (\ell^c\nabla_c f + \kappa f)g_{ab} \nonumber\\
\delta\phi &=& (\ell^c\nabla_c h + \kappa h) ,
\label{b7}
\end{eqnarray}
the Lagrangian \rref{b4} satisfies $\delta{\bf L}\sim\vartheta$.  We 
thus have an asymptotic symmetry in the sense described earlier.  In
particular, by restricting $f$ and $h$ to have their support in a small 
region near a horizon, we can make the variation $\delta I$ arbitrarily 
small.  For now, the relationship of $f$ and $h$ will remain unspecified; 
we shall see later that the choice that makes the transformation \rref{b7}
canonical actually implies that $\delta{\bf L}\sim\vartheta^2$.

Equation \rref{b7} is not enough to determine the separate variations
of $\ell^a$ and $n^a$.  This is to be expected, since the normalization
of $\ell^a$ is not fixed; the only restriction, from  \rref{b6}, is that
$n^a\nabla_a(n_b\delta\ell^b) = 0$.  We are thus free to choose 
$\delta\ell^a = 0$, which then implies that
\begin{eqnarray}
\ell_a\delta n^a &=& \ell^c\nabla_c f + \kappa f \nonumber\\
\delta\kappa &=& \ell^b\nabla_b(\ell^c\nabla_c f + \kappa f) \\
\delta s &=&  \ell^a\nabla_a(\ell^b\nabla_b h + \kappa h) \nonumber
\label{b8}
\end{eqnarray}
where $s = \ell^a\nabla_a\phi = \vartheta\phi$.  It follows that
\beq
[\delta_1,\delta_2]g_{ab} 
  = (\ell^c\nabla_c \{f_1,f_2\} + \kappa \{f_1,f_2\})g_{ab} \quad 
\hbox{with $\{f_1,f_2\} = (\ell^a\nabla_a f_1)f_2 - (\ell^a\nabla_a f_2)f_1$} ,
\label{b9}
\eeq
giving the standard conformal algebra.

To express the transformations \rref{b7} in Hamiltonian form, we need
the symplectic form $\Omega$ of \rref{a5}.  This can be computed by 
Wald's methods \cite{Wald,Kunstatter}.  For variations that have their
support only in a small neighborhood $\cal N$ of $\tilde\Delta$, the
main contribution will come from the integral along $\tilde\Delta$.
Restricting the symplectic form of Ref.\ \cite{Kunstatter} to $\tilde\Delta$, 
one finds that
\beq
\hat\Omega = {1\over2G}\int_{\tilde\Delta}
   \left( \ell^a\nabla_a(\delta_1\phi) \ell_b\delta_2 n^b
   - \ell^a\nabla_a(\delta_2\phi) \ell_b\delta_1 n^b \right){\hat\epsilon} ,
\label{b10}
\eeq
where $\hat\epsilon = n$ is the induced volume element on $\tilde\Delta$.  
Since $\tilde\Delta$ is null, one can integrate by parts, and use \rref{b8}
to obtain
\beq
\hat\Omega[\delta_1,\delta_2] = 
   -{1\over2G}\int_{\tilde\Delta} \left(\delta_1\phi\delta_2\kappa
   - \delta_2\phi\delta_1\kappa\right){\hat\epsilon} .
\label{b11}
\eeq

\section{Hamiltonian and Virasoro algebra}

The next question is whether the transformation \rref{b7} is canonical,
that is, whether it is generated by a ``Hamiltonian'' $L$.  Such a Hamiltonian
 must satisfy \cite{Wald}
\beq
\delta L[f,h] = \hat\Omega[\delta,\delta_{f,h}]
   = -{1\over2G}\int_{\tilde\Delta} \left[
   \delta\phi\, \ell^b\nabla_b(\ell^c\nabla_c f + \kappa f) 
   - \delta\kappa\, (\ell^c\nabla_c h + \kappa h) \right]{\hat\epsilon} 
\label{c1}
\eeq
where again $s=\ell^a\nabla_a\phi$ and I have integrated by parts 
to obtain the last equality.  The variation $\delta$ can be thought of
as an exterior derivative on the space of fields, and the integrability
condition for \rref{c1} is that $\delta^2L[f,h] = 0$.  If we assume that
the parameters $f$ and $h$ are field-independent, so $\delta f = 
\delta h = 0$, it is easy to see that this condition requires that $\delta s$
be proportional to $\delta\kappa$.  In particular, this proportionality
must hold for variations of the form \rref{b7}, and this, together with 
the requirement that $f$ and $h$ be field-independent, implies that 
\beq{\kappa\over s} = \hbox{constant on $\tilde\Delta$}  ,
\label{c2}
\eeq
\beq
sf = \kappa h .
\label{c3}
\eeq

Despite appearances, \rref{c2} is not a real restriction on the geometry, 
since it can always be satisfied by rescaling  $\ell^a$ as in \rref{b6a}.
As noted earlier, this relation makes the transformation \rref{b7} an 
even better approximate symmetry.  Indeed, it may be checked that now
\beq
\delta\int\phi R\epsilon = 2\int\left[ 
   n^a\nabla_af(\ell^a\nabla_a s - \kappa s) 
   + n^a\nabla_a\left({s\over\kappa}f\right)\ell^a\nabla_a\kappa
  \right] \epsilon ,
\label{c3a}
\eeq
and the integrand goes as $\vartheta^2$ near $\Delta$.  While no 
corresponding suppression appears automatically in the potential term 
in \rref{b4}, the variation of that term can easily be arranged to be of 
order $\vartheta^2$ by an appropriate choice of $\kappa/s$ on 
$\tilde\Delta$.

With the relation \rref{c3} between $h$ and $f$, \rref{c1} can be 
easily integrated, yielding
\beq
L[f] = {1\over2G}\int_{\tilde\Delta} s\left( 2\ell^a\nabla_a f + \kappa f
   \right) {\hat\epsilon}  = -{1\over2G}\int_{\tilde\Delta}\left(
   2\ell^a\nabla_as-\kappa s\right)f{\hat\epsilon} .
\label{c4}
\eeq
We must next choose a basis for the functions $f$ on $\tilde\Delta$.  
Since the normalization of $\ell^a$ is not fixed---even \rref{c2} 
determines it only up to a constant---the corresponding light 
cone coordinate has no intrinsic physical meaning.  There is,
however, a natural coordinate on $\tilde\Delta$, the dilaton $\phi$
itself, which by the two-dimensional version of the Raychaudhuri
equation should be monotonic on $\tilde\Delta$.  Let
\beq
z = e^{2\pi i\phi/\phi_+} ,
\label{c4a}
\eeq
where $\phi_+$ is the value of $\phi$ on the horizon, so $z\rightarrow 1$
at $\Delta$.\footnote{This choice is almost unique, in that $\phi_+$ is 
the only natural quantity in the theory having the right dimension.  In
principle, though, we could have chosen $w=z^\alpha$ to define our modes.
This would leave the central charge \rref{c9} unchanged, but would shift 
the Hamiltonian \rref{c7}.}  We can then choose a basis of functions to be 
proportional to $z^n$, with the proportionality constants determined from 
\rref{b9} and the requirement that the $f_n$ satisfy the standard $\hbox{Diff}
(S^1)$ commutation relations:
\beq
f_n =  {\phi_+\over2\pi s}z^n , \quad  \{f_m,f_n\} = i(m-n)f_{m+n} .
\label{c5}
\eeq
Note that with this choice, the consistency condition \rref{a2a} is satisfied  
asymptotically:  
\beq
(\ell^a\nabla_a)^2f_n = - {2\pi n^2\over\phi_+}sz^n \rightarrow0 \quad
\hbox{as $\tilde\Delta\rightarrow\Delta$} .
\label{c11}
\eeq
In terms of these modes, the Hamiltonian \rref{c4} becomes
\beq
L[f_n] = -{1\over2G}{\kappa\over s}\int_{\tilde\Delta} \left(
  1 + 2{\ell^a\ell^b\nabla_a\nabla_b\phi\over\kappa}\right)sf_n 
  {\phi_+\over2\pi i}{dz\over z} .
\label{c6}
\eeq
On shell, though, $\nabla_a\nabla_b\phi \propto g_{ab}$ \cite{Kunstatter}, 
so the last term in \rref{c6} vanishes, giving
\beq
L[f_n] = -{1\over2G}{\kappa\over s}{\phi_+{}^2\over2\pi }\delta_{n0} .
\label{c7}
\eeq

It remains for us to compute the Poisson brackets $\{L[f_m],L[f_n]\}$.
This is most easily done directly from equation \rref{c1}, and a
straightforward computation yields
\beq
\{L[f_m],L[f_n]\} = \delta_{f_m}L[f_n]
   = -{2\pi i\over G}{s\over\kappa}n^3\delta_{m+n,0} .
\label{c8}
\eeq
This may be recognized as the expression for a central term in the 
Virasoro algebra, with central charge
\beq
c = -{24\pi\over G}{s\over\kappa} .
\label{c9}
\eeq
Inserting \rref{c7} and \rref{c8} into the Cardy formula \rref{b2}, we
obtain a density of states
\beq
\log\rho(L_0) = {2\pi\phi_+\over G} ,
\label{c10}
\eeq
giving exactly the standard Bekenstein-Hawking entropy for the 
two-dimensional dilaton black hole \cite{Kunstatter}.

In contrast to previous work on Virasoro algebras at the horizon, this
derivation has the nice feature that the central charge \rref{c9} does
not depend on the particular black hole being considered.  The algebra
may therefore be viewed as a universal one, with different black holes
represented by different values \rref{c7} of $L_0$.

For simplicity, I have dealt only with two-dimensional black holes.  An
extension to higher dimensions would clearly be of interest.  As noted above,
though, higher-dimensional general relativity may be dimensionally reduced
in the manner of Kaluza and Klein to two-dimensional dilaton gravity coupled
with extra ``matter'' fields (see, for example, \cite{Yoon}).  It is fairly easy 
to see that these added terms cannot contribute to the classical central charge 
\rref{c9}, although they might give quantum corrections.  The algebra derived 
here is thus more universal than it might seem.

As also noted above, we should probably worry further about the making 
the notion of an ``asymptotic symmetry'' used here more rigorous.  It may be 
useful to exploit a generalization of the symmetry \rref{a1} that exists in the 
presence of a {\em conformal\/} Killing vector $\eta^a$,  
\beq
\nabla_a\eta_b + \nabla_b\eta_a = \kappa g_{ab} .
\label{c12}
\eeq
It is not hard to check that the transformation
\beq
g_{ab}\rightarrow \left( \eta^c\nabla_c f + {n-2\over2}\kappa f  \right)g_{ab}
\label{c13}
\eeq
leaves the Einstein-Hilbert action invariant provided that $f$ is chosen to satisfy
\beq
\int g^{ab}\nabla_a\kappa\nabla_bf \epsilon = 0 .
\label{c14}
\eeq
Moreover, if the original metric $g_{ab}$ admits a conformal Killing vector,
it is easily checked that the transformed metric does as well.  Maintaining
the condition \rref{c14} is  more complicated, but at least one solutions 
exists: if both $\kappa$ and $f$ are functions of a single null coordinate $v$, 
\rref{c14} holds automatically, and is preserved by \rref{c13}.  Work on 
understanding the implications of this extended symmetry is in progress.

\vspace{1.5ex}
\begin{flushleft}
\large\bf Acknowledgments
\end{flushleft}

I would like to thank Peter Beach Carlip, age $2{3\over4}$, for all the
``black hole information'' and ``quantum gravity'' he gave me during
this research.  This work was also supported in part by Department of 
Energy grant DE-FG03-91ER40674.

\end{document}